\documentclass[pdflatex, sn-vancouver-num]{sn-jnl}

\usepackage{graphicx}%
\usepackage{multirow}%
\usepackage{amsmath,amssymb,amsfonts}%
\usepackage{amsthm}%
\usepackage{mathrsfs}%
\usepackage{bm}
\usepackage[title]{appendix}%
\usepackage{xcolor}%
\usepackage{textcomp}%
\usepackage{manyfoot}%
\usepackage{booktabs}%
\usepackage{algpseudocode}%
\usepackage{listings}%
\usepackage{tabularx}%
\usepackage{wrapfig}%
\usepackage{bm}%
\usepackage{pifont}%

\theoremstyle{thmstyleone}%
\theoremstyle{thmstyletwo}%

\theoremstyle{thmstylethree}%

\begin{document}

\title[Medical Image Classification and Interpretability]{Interpretable Deep Learning Framework for Improved Disease Classification in Medical Imaging}


\author*[1]{\fnm{Jutika} \sur{Borah}}\email{borah\_jutika@gauhati.ac.in}




\author[1]{\fnm{Hidam Kumarjit} \sur{Singh}}\email{kumarjit\_hidam@gauhati.ac.in}


\affil[1]{\orgname{Gauhati University}, \orgaddress{\city{Guwahati}, \postcode{781014}, \country{India}}}





\abstract{Deep learning models have gained increasing adoption in medical image analysis. However, these models often produce overconfident predictions, which can compromise clinical accuracy and reliability. Bridging the gap between high-performance and awareness of uncertainty remains a crucial challenge in biomedical imaging applications. This study focuses on developing a unified deep learning framework for enhancing feature integration, interpretability, and reliability in prediction. We introduced a cross-guided channel spatial attention architecture that fuses feature representations extracted from EfficientNetB4 and ResNet34. Bidirectional attention approach enables the exchange of information across networks with differing receptive fields, enhancing discriminative and contextual feature learning. For quantitative predictive uncertainty assessment, Monte Carlo (MC)-Dropout is integrated with conformal prediction. This provides statistically valid prediction sets with entropy-based uncertainty visualization. The framework is evaluated on four medical imaging benchmark datasets: chest X-rays of COVID-19, Tuberculosis, Pneumonia, and retinal Optical Coherence Tomography (OCT) images. The proposed framework achieved strong classification performance with an AUC of 99.75\% for COVID-19, 100\% for Tuberculosis, 99.3\% for Pneumonia chest X-rays, and 98.69\% for retinal OCT images. Uncertainty-aware inference yields calibrated prediction sets with interpretable examples of uncertainty, showing transparency. The results demonstrate that bidirectional cross-attention with uncertainty quantification can improve performance and transparency in medical image classification.}



\maketitle

\section{Introduction}

Classification is a crucial process in the interpretation of medical images, supporting clinical diagnosis and treatment planning. However, accurate classification remains critical, as errors can lead to misdiagnosis and delayed treatment. Deep Learning (DL)-based lung disorder detection faces notable challenges, including false positives, false negatives, and radiological image complexities such as overlapping abnormalities, intricate anatomical structures, and subtle disease patterns \cite{rajpurkar2017chexnet}, \cite{he2025comparative}. Standard Convolutional Neural Networks (CNNs) often fail to address these challenges due to their limited ability to capture global context, multiscale features representation, and contextual relationships, which are crucial for accurate diagnosis \cite{9156697}, \cite{ramamurthy2024integration}. Attention mechanisms have emerged as a powerful tool in DL, improving model robustness and generalizability by enabling models to prioritize the most informative features dynamically and enhancing adaptability to changes in data distribution \cite{oktay2018attention}, \cite{abbasi2025enhanced}.

In medical image analysis, where data variability and complexity pose significant challenges, attention-based feature fusion proves to be particularly advantageous \cite{zheng2023casf}. Recent studies \cite{valanarasu2021medical}, \cite{li2022transbtsv2} have explored transformer-based approaches; however, self-attention mechanisms often struggle to capture fine-grained local features, leading to difficulties in distinguishing objects of interest from complex backgrounds.  Recent advancements, such as Medical Transformer \cite{valanarasu2021medical}, GasHis Transformer \cite{chen2022gashis}, have demonstrated the efficacy of combining CNNs with transformers to capture both global and local features for tasks like disease detection and segmentation. These approaches emphasize the simultaneous learning of local spatial structures and long-range dependencies to reduce noise and improve feature representation. They offer potential for scalable and early diagnosis. However, these models exhibit high computational cost, making them impractical in resource-constrained areas. 

\begin{figure}[h!]
    \centering
    \includegraphics[width=0.85\linewidth]{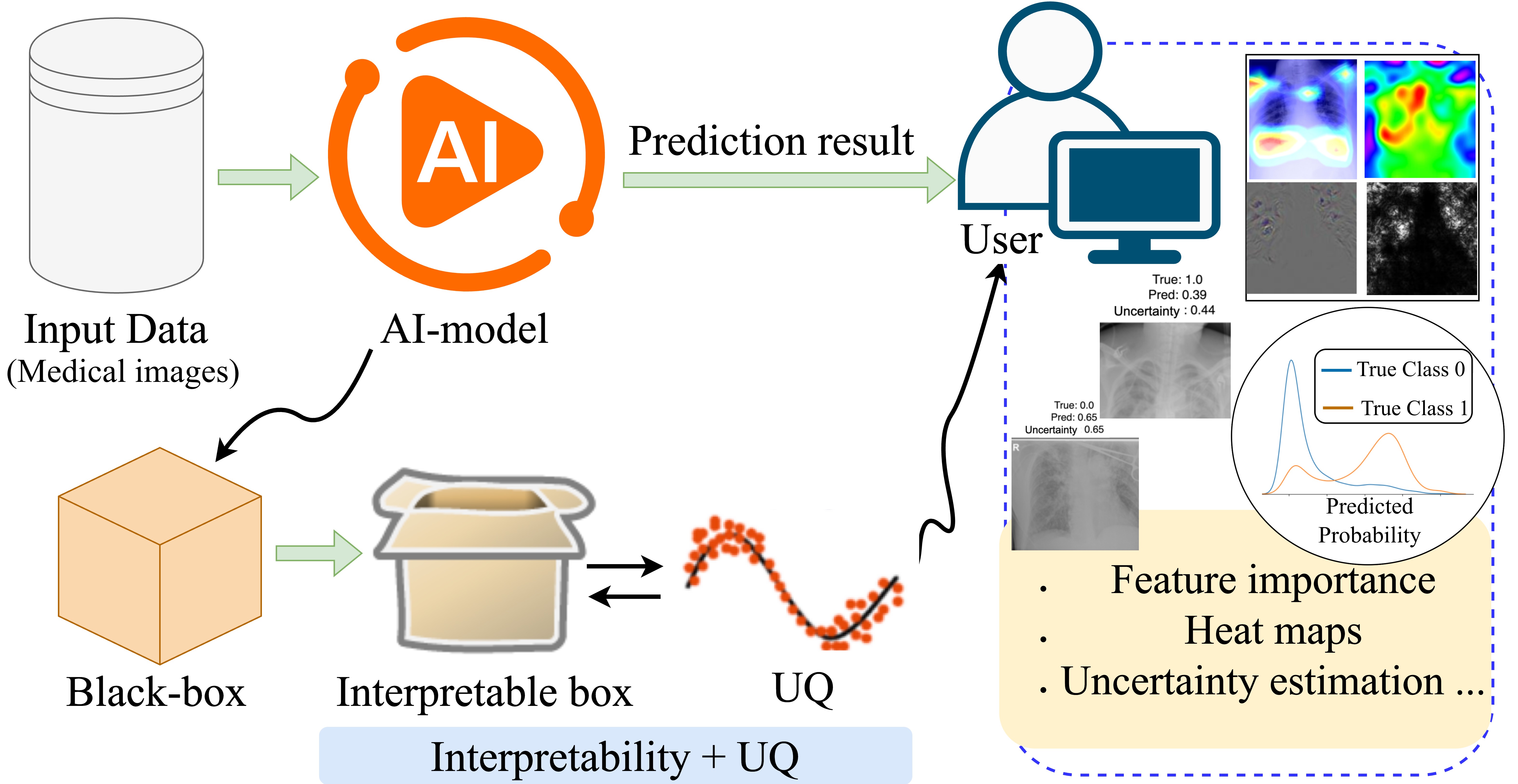}
    \caption{Overview of a typical AI framework with interpretable methods. Conceptual schematic illustrating the role of explainable and interpretable AI in clinical decision-making.}
    \label{fig1}
\end{figure}

Unlike traditional fusion methods like concatenation or element-wise addition, cross-attention captures complex dependencies and dynamically weights contributions from each modality, resulting in more informative and nuanced representations. Furthermore, clinical diagnosis requires interpretable models that support clinical decision-making. The inherent complexity of DL models often limits transparency, making it difficult for clinicians to understand how decisions are made. Class imbalance in datasets, where rare diseases are underrepresented, also leads to poor model generalization and biased predictions.  Uncertainty Quantification (UQ) is critical to identify low-confidence predictions, enabling clinicians to prioritize cases for manual review. Thus, interpretability and UQ are pivotal challenges \cite{tjoa2020survey} \cite{doshi2025deep} in the application of DL models to radiological image classification \cite{faghani2023quantifying}, \cite{huang2024review}. This is particularly problematic in radiology, where high inter-patient variability, differences in imaging protocols, and noise can lead to inconsistent classification outcomes \cite{kinger2024review}. Moreover, traditional classification networks frequently fail to provide uncertainty estimates, undermining their reliability in clinical practice. Therefore, quantifying uncertainty alongside validity, such as coverage guarantees, is crucial for assessing the confidence in model predictions as shown in Figure~\ref{fig1}.

Our work introduces a dual cross-attention hierarchical feature fusion framework with an enhanced convolution block attention module. By integrating cross-attention between the EfficientNetB4 and ResNet34 architectures, our model effectively captures multi-scale features, preserving both low-level spatial textures and high-level semantic representations. This approach leverages attention-guided multi-scale global-local fusion to enhance feature representation. Validated on four medical imaging datasets, our proposed attention-based framework demonstrates superior performance in the classification of lung diseases from chest X-ray images and retinal diseases from OCT images. Typically, for automated classification, softmax outputs are optimized for model performance, which are far from conveying the true certainty of a disease.  Since real-world radiology holds a lot of uncertainty, softmax probabilities alone are not representative of calibrated probabilities. This work introduces conformal prediction (CP) on top of the calibrated predictive distribution of MC-Dropout, adding an extra level of interpretability and reliability. This work sets a formal coverage guarantee with predictions, ensuring the true class label is contained within the predicted set with a specified confidence level. Experimental results in four publicly available datasets demonstrate that the proposed framework achieves strong classification performance. The primary contributions are summarized as follows:

\begin{enumerate}
    \item We introduce a cross-attention mechanism that dynamically fuses feature maps from two efficient networks, EfficientNetB4 and ResNet34. The framework leverages the strength of both architectures, bidirectional feature exchange, significantly improving feature representation for accurate medical image classification. 

    \item Incorporating UQ together with interpretability methods provides confidence estimates for each prediction, demonstrating reliability with uncertainty measures pointing towards trustworthiness in clinical decision making.

    \item We used a conformal prediction, a model-agnostic and distribution-independent method that expresses uncertainty not only as a scalar metric but also as guaranteed, interpretable output.
\end{enumerate}

The remaining paper follows the given hierarchy: related works in Section \ref{sec2}, while the details of the proposed framework are provided in Section \ref{sec3}. Experimental details are provided in Section \ref{sec4}, followed by the results and analysis in Section \ref{sec5}. Section \ref{sec6} presents an in-depth discussion from both DL and clinical perspectives. Finally, Section \ref{sec7} mentions the conclusion and future work.

\section{Related Works}\label{sec2}

Attention mechanisms have emerged as a powerful tool in computer vision and can be broadly categorized into channel attention \cite{hu2018squeeze}, spatial attention \cite{woo2018cbam}, and cross attention, each addressing specific aspects of feature representation. Cross-attention mechanisms have gained traction for their ability to model relationships between modalities or feature maps. Unlike self-attention, which operates within a single feature space, cross-attention enables exchange between distinct feature representations, enhancing the model's ability to capture contextual dependencies. For instance, authors in \cite{schlemper2019attention} introduced an attention gate model for medical image analysis that suppresses irrelevant regions while highlighting task-specific features. Similarly, authors in \cite{guan2020thorax} proposed a CNN with three branches of attention for identifying thorax diseases, integrating global and local signals to enhance performance. These studies underscore the potential of attention mechanisms to address challenges such as noise suppression and feature refinement. Despite advancements in attention mechanisms, several limitations persist, particularly in cross-attention multi-network feature fusion. Although channel and spatial attention mechanisms have been studied extensively, their integration with cross-attention has not been explored sufficiently. Finally, the lack of UQ in attention limits their reliability in medical imaging like critical applications.  

UQ techniques have gained significant attention in the last couple of years in medical data analysis \cite{faghani2023quantifying}, \cite{fayjie2024predictive}. Identifying uncertain samples can help flag complex cases for further investigation, rather than making overconfident, incorrect predictions. In \cite{gawlikowski2023survey}, factors that cause uncertainty are highlighted, including variability in real-world situations, errors in measurement systems, model structure, training data, and unknown data. Categorized as aleatoric and epistemic uncertainty \cite{kendall2017uncertainties}, aleatoric uncertainty arises from inherent noise within the data, while epistemic uncertainty stems from model limitations. In \cite{gautam2024fusionexnet}, the authors introduced an interpretable DL framework for the detection of skin cancer, combining pre-trained networks EfficientNetV2S and XceptionNet for the enhancement of feature representation. They included faster score-CAM for interpretability to support clinical interpretability. In \cite{mecili2024efficient}, the authors proposed an explainable DNN framework for the detection of diabetes. They integrated hybrid feature extraction with advanced data augmentation to improve performance on classification and interpretability. The authors in \cite{nair2020exploring} demonstrated the utility of MC-dropout in the detection and segmentation of multiple sclerosis, showing improved performance and uncertainty estimation. In \cite{chagas2023uncertainty}, the authors applied MC-Dropout and test-time augmentation for classifying membranous nephropathy, showing the importance of correlation between prediction variance and accuracy. While ensemble learning methods \cite{dutschmann2023large}, \cite{fayjie2024predictive} have advanced UQ, they are computationally expensive in resource-constrained environments. The reliability of uncertainty quantification depends on the model and data distribution. The traditional calibration of the softmax scores is the most widely used method. Few studies by \cite{romano2020classification}, \cite{hechtlinger2018cautious} introduced methods for equal coverage for each class. However, these methods have limitations from empirical evaluations of the state-of-the-art. Concretely, empirical evaluation of CP with uncertainty quantification for radiological medical image classification also remains restricted and underexplored.

\begin{figure}[h!]
 \centering
        \includegraphics[width=0.95\textwidth]{smaple_class_distribution.jpg}
 \caption{(a) Random example images from the four datasets. From top: Tuberculosis chest X-ray, Covid Chest X-ray, Pneumonia Chest X-ray, Retinal OCT images, (b) Visualization of class distribution for each dataset. The random samples illustrate the visual variability in contrast, appearance, and anatomical structures in the data.}
\label{fig2}
\end{figure}

\section{Materials and Methods}\label{sec3}
In our classification task, we trained and evaluated the proposed method using four publicly available medical imaging datasets: COVID-19 \cite{chowdhury2020can}, Tuberculosis \cite{rahman2020reliable}, Pneumonia \cite{kermany2018labeled} chest X-ray, and OCT \cite{kermany2018labeled} images. Figure~\ref{fig2}(a) and (b) show some randomly selected samples from four datasets and their class-wise distribution, respectively. We randomly split the data into two main sets, where 80\% of the data was used for training, and the remaining 20\% was used as a test set.

\subsection{Proposed Methodology}
The proposed framework is illustrated in Figure~\ref{fig3}. We used EfficientNetB4 and ResNet34 due to their complementary strengths in feature representations and performance efficiency trade-offs. EfficientNetB4, part of the Efficient family introduced in \cite{tan2019efficientnet} in 2019, employs a compound scaling method that uniformly scales depth, width, and resolution to optimize performance with fewer parameters. This scaling enables the network to capture fine-grained details and high-level semantic information. ResNet34, part of the Residual Network (ResNet) family introduced in \cite{he2016deep}, addresses the vanishing gradient problem through its residual learning framework while enabling the training of deep networks. The architecture extracts hierarchical features, with early layers capturing low-level details and deeper layers capturing high-level semantics. Its residual connections make it robust to over-fitting, even with limited training data. This contrast between the two architectures allows the proposed dual cross-attention module to exchange and integrate feature information across the two branches, thus improving representational diversity. 

\begin{figure}[h!]
    \centering
    \includegraphics[width=0.98\linewidth]{Cross_attention_framework.jpg}
    \caption{Conceptual overview of the proposed framework illustrating dual backbone feature extraction via EfficientNetB4 and ResNet34, a hierarchical cross-attention fusion mechanism, an uncertainty quantification module, and interpretability with CP.}
    \label{fig3}
\end{figure}

\subsubsection{Hierarchical Multi-Scale Feature Extraction}
\label{subsec:multi_scale}
We consider a medical image classification problem where each sample consists of an input image ($x$) and an associated label $y \in {1, 2, \dots, N_{cl}}$, with $N_{cl}$ representing the total number of classes. Feature maps extracted from specific layers of EfficientNetB4 and ResNet34 are combined to form a multi-scale representation.

The EfficientNetB4 network processes the input image $x$ through a series of Convolutional blocks. Feature maps extracted from EfficientNetB4 Block 3 and 4 have spatial sizes of $28 \times 28$ and $14 \times 14$, respectively. Similarly, the ResNet34 network processes the input through a series of residual layers, with feature maps extracted from ResNet Layers 2 and 3 having spatial sizes of $28 \times 28$ and $14 \times 14$, respectively. From the corresponding depth of both networks, multi-scale features are extracted to preserve scale consistency. These feature maps are denoted as, \[X^{eff}_{1} = \text{EfficientNetB4 Block 3(x)}, X^{eff}_{2} = \text{EfficientNetB4 Block 4(x)}\] \[X^{r}_{1} = \text{ResNet34 Layer 2(x)}, X^{r}_{2} = \text{ResNet34 Layer 3(x)}\]

The multi-scale feature extraction produces two sets of concatenated feature maps, \[F_{1} = [ X^{eff}_{1}, X^r_{1}] \in \mathbb{R}^{2C_{1} \times 28 \times 28}\] \[F_{2}= [ X^{eff}_{2}, X^r_{2}] \in \mathbb{R}^{2C_{1} \times 14 \times 14}\] where $F_{1}$ represents fine-grained features at a spatial resolution of $28 \times 28$, and $F_{2}$ represents high-level features at a spatial resolution of $14 \times 14$. Subscript $1$ and $2$ represent two hierarchical feature scales corresponding to intermediate and deeper layers of both the networks. Rather than fusing these concatenated features directly, the proposed framework forwards the scale-aligned feature maps from each backbone to a cross-attention fusion module. 

\subsubsection{Cross Attention mechanism}
Building on prior works such as \cite{he2020cabnet}, \cite{xie2021cross}, which proposed channel and spatial attention mechanisms to explore discriminative features, our module ensures the network prioritizes specific features via cross-attention, refining context-aware attention refinement. Traditional CBAM \cite{woo2018cbam} and SENet \cite{hu2018squeeze} applied sequential channel and spatial attention within a single network; our proposed framework explicitly models bi-directional feature interaction across the network via cross-attention. 

Let $X^{eff}_{img} \in \mathbb{R}^{C \times H \times W}$ and $X^{r}_{img} \in \mathbb{R}^{C \times H \times W}$ represent the scale-aligned feature maps from EfficientNetB4 and ResNet34 respectively corresponding to intermediate and high level semantic representations. To model inter-network dependencies, a bi-directional cross-attention fusion mechanism is employed independently; we compute $\text{Query}(Q), \text{Key}(K)$, and $\text{Value}(V)$ representations of both networks.

For EfficientNetB4, Q, termed as $Q_{eff}$, is derived from EfficientNetB4 features, while K and Q representations $K_{r}$ and $V_{r}$ are derived from ResNet34 features. Similarly, for ResNet34, $Q_{r}$ is derived from ResNet34 features, while K and Q representations $K_{eff}$ and $V_{eff}$ are derived from EfficientNetB4 features, respectively.

The cross-attention weights are computed using the scaled dot-product attention mechanism. For EfficientNetB4, the attention is computed as,

\begin{equation}
    \text{Attention}(Q_{eff}, K_{r}, V_{r})_{X^{eff}_{img}} = \text{Softmax} \left(\frac{Q_{eff}K_{r}^T}{\sqrt{d_k}} \right) V_{r}     
\end{equation}
Similarly, for ResNet34, the attention is computed as,
\begin{equation}
    \text{Attention}(Q_{r}, K_{eff}, V_{eff})_{X^{r}_{img}} = \text{Softmax} \left(\frac{Q_{r}K_{eff}^T}{\sqrt{d_k}} \right) V_{eff}
\end{equation}
where $Q_{eff}K_{r}^T$ and $Q_{r}K_{eff}^T$ measure cross-network feature affinity (similarity between queries and keys), and $\sqrt{d_k}$ is a scaling factor that stabilizes gradients during training. The final fused feature representation $F_{fusion}$ is obtained by aggregating the bi-directional attention output:
\begin{equation}
    F_{fusion} = \text{Attention}(Q_{eff},K_{r},V_{r}) + \text{Attention}(K_{eff},Q_{r},V_{eff})
\end{equation}

This bidirectional fusion enables both networks to mutually incorporate complementary information, capturing complementary patterns and contextual information across networks.

\subsubsection{Feature Refinement and Classification}
Following the cross-attention fusion, the fused feature map $F_{fusion}$ is refined further using a modified channel-spatial attention mechanism inspired by CBAM. This attention mechanism extends the standard CBAM by incorporating cross-network contextual feature extraction. Unlike standard CBAM, the proposed framework refinement operates on cross-attention fused features, providing inter-network contextual cues.

Channel attention is applied to emphasize informative feature channels. Global average pooling and max pooling is first applied to the fused feature representations, 

\[ z_c = [\text{Avg}(F_{fusion}); \text{Max}(F_{fusion})]\]

The pooled representations are passed through a shared multi-layer perceptron (MLP) to generate channel-wise attention weights: 
\[M_{c} = \sigma((W_2 \delta (W_1z_c))\] 

where $W_1$ and $W_2$ are shared learnable weights, $\delta(\cdot)$ is ReLU activation, and $\sigma(\cdot)$ is the sigmoid activation function. 

Similarly, for the spatial attention mechanism, attention maps are computed for each spatial location in the fused feature maps, given as, 

\[z_{s} = [\text{Avg}_c(F_{fusion}); \text{Max}_c(F_{fusion})]\]. 

Spatial feature map is generated as:

\[M_{s} = \sigma(f^{7 \times 7}(z_s))\] 

where $f^{7 \times 7}$ denotes a $7 \times 7$ convolutional kernel. The  final refined feature is obtained as,

\[F_{final} = M_s \odot (M_c \odot F_{\text{fusion}})\] where $\odot$ represent element-wise multiplication.

The final refined feature map is passed through an adaptive average pooling layer to produce fixed size representation, ensuring compatibility with a fully connected layer regardless of the input size.

\begin{equation}
    F_{pooled} = \text{AdaptiveAvgPool}(F_{final})
\end{equation}

The pooled features are then flattened into a vector and passed through a fully connected layer with weights $W_{c}$ and bias $b$,

\begin{equation}
    y = \text{Softmax}(W_{c}F_{pooled} + b)
\end{equation}

Here, $y$ represents the probability distribution of output over possible class labels. The Softmax activation provides the probabilistic interpretation of the model's predictions. The class with the highest probability is selected as the predicted label $\hat{y}$. 

\subsection{Model Uncertainty Calculation}
\subsubsection{MC-Dropout} 
To further quantify predictive reliability, MC-Dropout is applied during inference. A widely used technique for quantifying uncertainty introduced by \cite{gal2016dropout} treats dropout regularization as a Bayesian approximation. This approach involves running multiple stochastic forward passes on the same input, effectively generating a predictive distribution. The resulting Softmax outputs are averaged to form the final prediction while also reflecting the uncertainty in the model's prediction.  
Consider a trained network model, $f_\theta(\cdot)$, where $f$ is the network and $\theta$ are the learnable parameters with active dropout. Suppose a new test {$x, y$} is to be predicted, where $x$ is the image array and $y$ is the corresponding label. It will be passed through the network with active dropout $M$ times (i.e., the number of forward passes). Thus, the predictive posterior distribution can be approximated to generate the final prediction as follows,

\begin{equation}
    p_{\theta}(y|x) =\frac{1}{M}\sum_{m=1}^{M} {p}_m 
\end{equation}
where ${p}_m$ denotes the output of the softmax function at pass $m \in$ {$1,..., M$}.

The mean predictive probability of MC-Dropout is used to compute $p_\theta(y|x)$, and the entropy values are used to assess the epistemic uncertainty. We computed the posterior probability distribution over the trained network weights, given the input samples and the corresponding ground truths, with a dropout rate of 0.5 and 100 forward passes.

\subsection{Conformal Prediction}
To obtain valid confidence estimates for classification, we integrated inductive conformal prediction (ICP) into our proposed framework. Samples with higher predictive entropy under MC-Dropout resulted in larger conformal prediction sets, indicating a positive association between the model uncertainty and the prediction set coverage size. The datasets were randomly split into three disjoint subsets comprising 70\% for training, 10\% for calibration, and 20\% for testing with strict subject-level separation across splits. The model was trained exclusively on the training set while non-conformity scores, measuring the disagreement between model predictions and ground-truth labels, were computed using calibration sets. The combination of MC-Dropout with CP provides enables both probabilistic uncertainty quantification and formal coverage guarantees, offering interpretable and reliable predictions.

ICP was employed for both binary and multiclass classification tasks to generate prediction sets containing the true label with a user-specified probability. In binary classification, the nonconformity score for the model output probability $p \in [0, 1]$ of the model is defined as, 
\[s = \begin{cases}
    1 - p, & \text{if true class label} = 1\\
    p, & \text{if true class label} = 0
\end{cases}\]

For multiclass classification, the output probability vector $p = (p_{1}, p_{2},....,p_{k})$ of the model where $K = 4$ in our case. Nonconformity score is defined as $s = 1 - p_{y}$ reflecting the model uncertainty in the true class probability predicted. A separate calibration set is used for the computation of nonconformity scores. $\tau$ a conformal threshold was selected as $(1-\alpha)$-quantile of these scores, where $\tau = \text{quantile}_{1-\alpha}(S_{val})$. Here, we considered $\alpha = 0.1$, corresponding to a 90\% target confidence level. The prediction set $\Gamma(x)$ is given by, 

\[\Gamma(x) = \{c \in \{1, \ldots, K\} : 1 - p_{c} \leq \tau\}\] 

We used a fallback rule to avoid empty prediction sets. If no class satisfies this condition ($\Gamma(x) = \phi$), a fallback rule assigns the most probable class $\hat{y} = \text{argmax}_y p_\theta(y|x)$. This guarantees that each prediction is non-empty while preserving the theoretical coverage property of CP.

\section{Experiments}\label{sec4}
\label{exp}
\subsection{Parameter Settings} 
We implement the proposed mechanism and perform experiments using the PyTorch library. Training and evaluations have been accomplished using the NVIDIA GeForce RTX 3060 Ti built-in GPU. The batch size for training is set to 32. The maximum epochs is set to 200, and the learning rate is initialized to $3e-4$ with a weight decay of $1e-4$. We chose Adam as the optimizer for the training stage. The whole training process takes 2h30min, and the testing time for a radiology image is approximately 0.015s.

\subsubsection{Evaluation metrics}
For a comprehensive overview and evaluation of the performance of the proposed method, we employed standard performance metrics like accuracy (Acc), AUC is the area under the ROC curve (AUROC), Area Under the Precision-Recall (AUPR), precision (Pre), recall (Rec), specificity (Sp), F1-score, Matthews Correlation Coefficient (MCC) and Cohen's Kappa (CK).  As our dataset is class-imbalanced, we use the AUPR as a metric that evaluates classification performance. AUC and AUPR focus on positive class performance. For uncertainty quantification, entropy looks into the complexity of the distribution and how it compares to a uniform distribution. It's a measure of variability associated with random variables \cite{shannon1948mathematical}.  When high entropy signals the model is uncertain and might benefit from additional information, and vice versa. Since medical imaging deals with complex, noisy, and ambiguous data, images like X-rays can have varying levels of quality and precision. In the context of predictive uncertainty, entropy measures the uncertainty of a probability distribution over possible outcomes and is defined mathematically as - 

\begin{equation}
    H(p(y | x)) = - \sum_{c=1}^{C} p(y | x)log p(y | x)
\end{equation}
where C is the number of classes. For un

\begin{table}[htbp]
\centering
\caption{Performance Comparison of our proposed method applied to four datasets during the evaluation stage with uncertainty quantification. HUS $\rightarrow$ High Uncertainty Sample}\label{tab1}
\begin{tabular}{lccccc}
\toprule
$\bm{\textbf{Datasets}}$ & $\bm{\textbf{Acc}}$ & $\bm{\textbf{Mean Entropy}}$ & $\bm{\textbf{Standard Deviation}}$ & $\bm{\textbf{HUS}}$ & $\bm{\textbf{Misclassified Samples}}$\\
\midrule
Covid-19  & \textbf{97.41} & 0.0254  & 0.1011  & 3 & 12\\

Tuberculosis  & \textbf{99.21} & 0.0024 & 0.0353  & 7 & 10\\

Pneumonia  & \textbf{98.95} & 0.0150 & 0.0817 & 5  & 11\\

OCT  & \textbf{91.17}  & 0.1603 & 0.2457 &  45 & 36\\
\bottomrule
\end{tabular}
\end{table}

\begin{table}[htbp]
\centering
\caption{Details of class-wise classification performance of four datasets used in this study. CNV
$\rightarrow$ Choroidal Neovascularization, DME $\rightarrow$ Diabetic Macular Edema}\label{tab2}%
\begin{tabular}{@{}llllll@{}}
\toprule
$\bm{\textbf{Datasets}}$ & $\bm{\textbf{Classes}}$ & $\bm{\textbf{Pre}}$  & $\bm{\textbf{Rec}}$ & $\bm{\textbf{Sp}}$ & $\bm{\textbf{F1-score}}$\\
\midrule
Covid-19 Chest X-ray    & Covid-19   & 94.63  & 98.17 & 98.17 & 96.37\\
    &  Lung Opacity  &  98.86 & 96.60 & 96.53 & 97.72\\

Pneumonia Chest X-ray     & Pneumonia  &  98.29  & 98.55  &  98. 63 & 98.42\\
    &  Normal & 96.14  & 95.47 & 95.43 & 95.80\\

Tuberculosis Chest X-ray    & Tuberculosis   &  99.21 & 94.74 & 94.74 & 96.92\\
    & Normal  & 96.02  & 99.41 & 99.41  & 97.68\\

Retinal OCT & CNV   &  92.90 & 90.04 & 97.39 & 91.45\\
    &  DME  &  96.13 & 91.24 & 98.80 & 93.62\\
    &  DRUSEN  &  88.59 & 83.87 & 96.98 & 86.16\\
    &  NORMAL &  87.44 & 98.43 & 95.04 & 92.61\\

\hline
\end{tabular}
\end{table}

\begin{table*}[!h]
\caption{Performance comparison of our proposed DCAT model applied to various chest X-ray and retinal OCT images during the evaluation stage}\label{tab3}
\begin{tabular}{@{\extracolsep\fill}lcccccccc}
\toprule%
\textbf{Datasets} & \textbf{AUC} & \textbf{AUPR} & \textbf{Pre} & \textbf{Rec} & \textbf{Sp} & \shortstack{\textbf{F1-Score}} & \textbf{MCC} & \textbf{CK}\\
\midrule
Covid-19  & 99.75 & 99.81 & 98.21  & 97.35 & 97.50 & 97.78 & 94.68 & 94.67\\

Tuberculosis  & 100.0 & 100.0 & 100.0 & 98.68 & 100.0 & 99.34 & 98.83 & 98.82\\

Pneumonia  & 99.93 & 99.97  & 99.21 & 99.34 & 97.91 & 99.27 & 97.35 & 97.35 \\

OCT & 98.69 & 96.36 & 91.33 & 91.17 & 97.05 & 91.13 & 88.27 & 88.17\\
\hline
\end{tabular}
\end{table*} 

\begin{figure}[h!]
    \centering
    \includegraphics[width=0.98\linewidth]{AUROC_Curves.jpg}
    \caption{AUROC Curves for all four datasets as part of our evaluation framework. Curves with multiple classes illustrating the trend of true positive rate versus false positive rate.}
    \label{fig4}
\end{figure}

\section{Results and Discussion}\label{sec5}
A comprehensive evaluation of the performance of the proposed model was conducted, and the findings are reported through performance metrics in the subsequent tables. To evaluate the effectiveness of the proposed model, we experimented on four different medical imaging datasets as listed in the earlier section. The experimental results of the four datasets are shown in Table~\ref{tab1}, \ref{tab2}, and \ref{tab3}, respectively. The results of our ablation study, shown in Table~\ref{tab4}, effectively demonstrate the performance of various individual components in classification performance.  We also evaluated and made a comparison study with some of the already published works. Table~\ref{tab5} demonstrates the effectiveness of our proposed framework. It achieved an AUC of 99.75\%, 100\%, 99.93\%, and 98.69\% on COVID-19 chest X-ray, tuberculosis chest X-ray, pneumonia chest X-ray, and retinal OCT datasets, respectively. The AUROC and PR-curves are shown in Figures~\ref{fig4} and \ref{fig5}, for the three chest X-ray datasets and the retinal OCT dataset, respectively. The CP evaluation results in Table~\ref{tab6} suggest that the prediction sets produced by the conformal predictor are informative both for the binary and multiclass datasets, achieving near-nominal coverage. 
High entropy in MC-Dropout indicates greater epistemic uncertainty in the model, while larger CP sets indicate lower confidence in assigning a unique label at a fixed confidence interval. Significantly, the HUS given in Table \ref{tab1} consistently corresponds to instances where the conformal predictor yielded multi-class prediction sets $\Gamma(x) > 1$. These results validate the coherence of the proposed reliability framework, where MC Dropout quantifies model-level uncertainty, and CP effectively translates this uncertainty into a formal coverage guarantee.

\begin{figure}[h!]
    \centering
    \includegraphics[width=0.98\linewidth]{PR_Curves.jpg}
    \caption{PR Curves for all four datasets as part of our evaluation framework. Curves with multiple classes illustrating the trend of precision versus recall.}
    \label{fig5}
\end{figure}

\subsection{Ablation Study}
A comprehensive ablation study was conducted to validate and justify the component of the proposed model in the four data sets and the evaluation metrics previously discussed. The analysis supports our design choices, providing valuable insights into the significance of each element in enhancing performance and achieving the desired outcomes. Table~\ref{tab4} summarizes the results of the datasets without the proposed framework. This shows that each component of the model contributes to the overall performance. Table~\ref{tab5} compares our proposed method with other recently proposed deep learning methods.

\begin{table}[h!]
\centering
\caption{Performance comparison of EfficientNetB4 and ResNet34 at an individual level without cross-attention mechanism on four datasets}\label{tab4}
\begin{tabular}{lcccccc}
\toprule
\multirow{2}{*}{\textbf{Datasets}} & 
\multicolumn{5}{c}{\textbf{Individual model performance}} \\
\cmidrule(lr){2-7}
 & \multicolumn{2}{c}{\textbf{ResNet34}} & \multicolumn{3}{c}{\textbf{EfficientNetB4}} \\
\cmidrule(lr){2-4} \cmidrule(lr){5-7}
 & $\bm{\textup{Acc}}$ & $\bm{\textup{Entropy}}$ & $\bm{\textup{Std. Dev.}}$ & $\bm{\textup{Acc}}$ & $\bm{\textup{Entropy}}$ & $\bm{\textup{Std. Dev.}}$ \\
\midrule
Covid-19  & 86.99 & 0.0443 & 0.0239 &  88.76 & 0.0926 & 0.0172\\

Tuberculosis  & 80.94 & 0.1896  & 0.0948  & 85.89 & 0.2000 & 0.0968\\

Pneumonia  & 81.89 & 0.0500  & 0.0467  & 80.77 & 0.0756 & 0.0891\\

OCT  & 88.91 & 0.2646  & 0.0367 & 90.18 & 0.1290 & 0.2567\\
\bottomrule
\end{tabular}
\end{table}

\begin{figure}[!h]
    \centering
    \includegraphics[width=.98\textwidth]{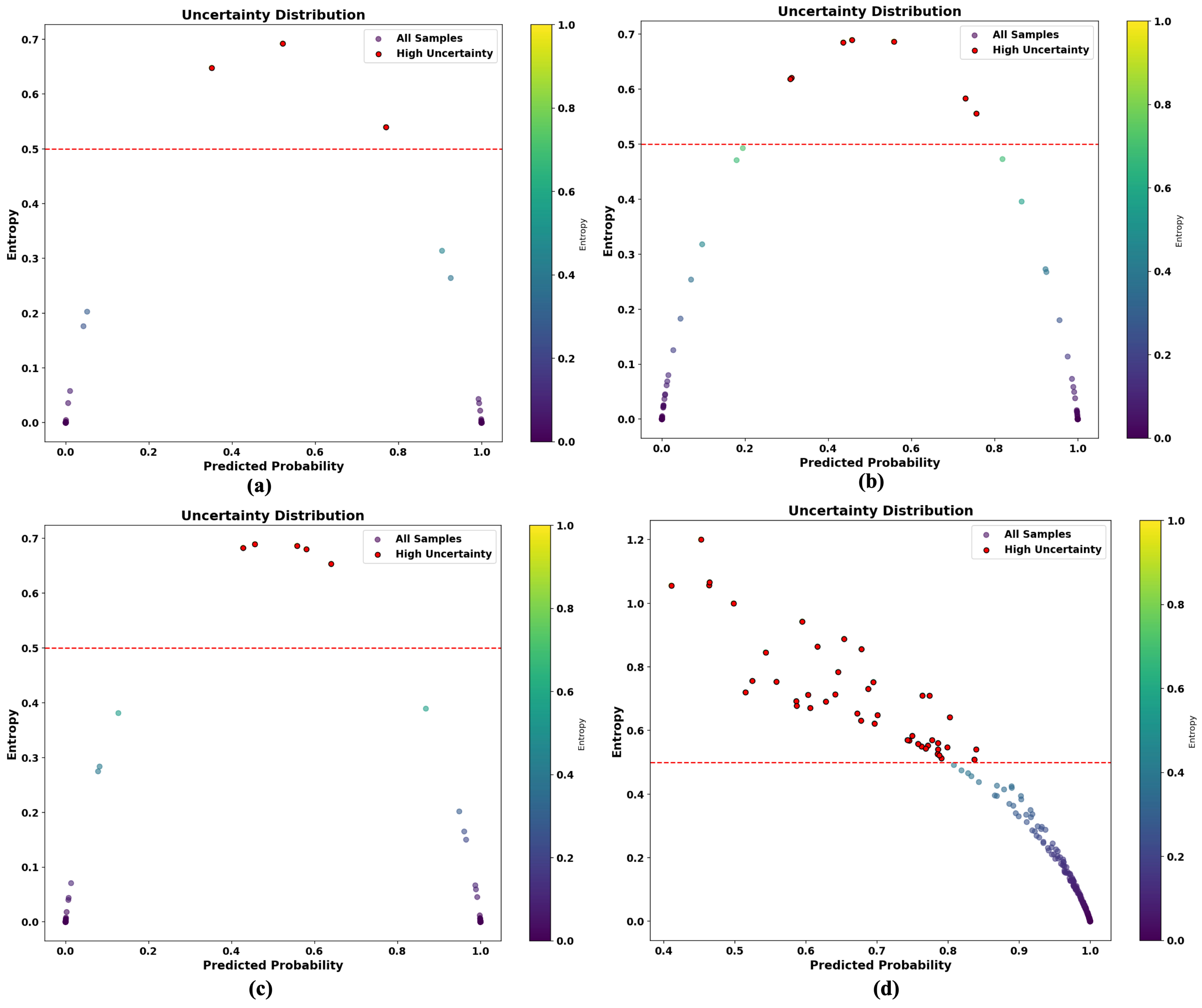}
    \caption{Uncertainty distribution visualization of a subset of data from each dataset above an uncertainty threshold. From the left: (a) Covid-19 Chest X-ray, (b) Tuberculosis Chest X-ray, (c) Pneumonia Chest X-ray, and (d) Retinal OCT. Scatter plot illustrating the variation of entropy versus predicted probability for a test subset of data with scattered points distributed across the range of probability. The horizontal dashed line indicate threshold for high-uncertainty samples.}\label{fig6}
\end{figure}

\begin{figure}[!h]
    \centering
    \includegraphics[width=0.98\textwidth]{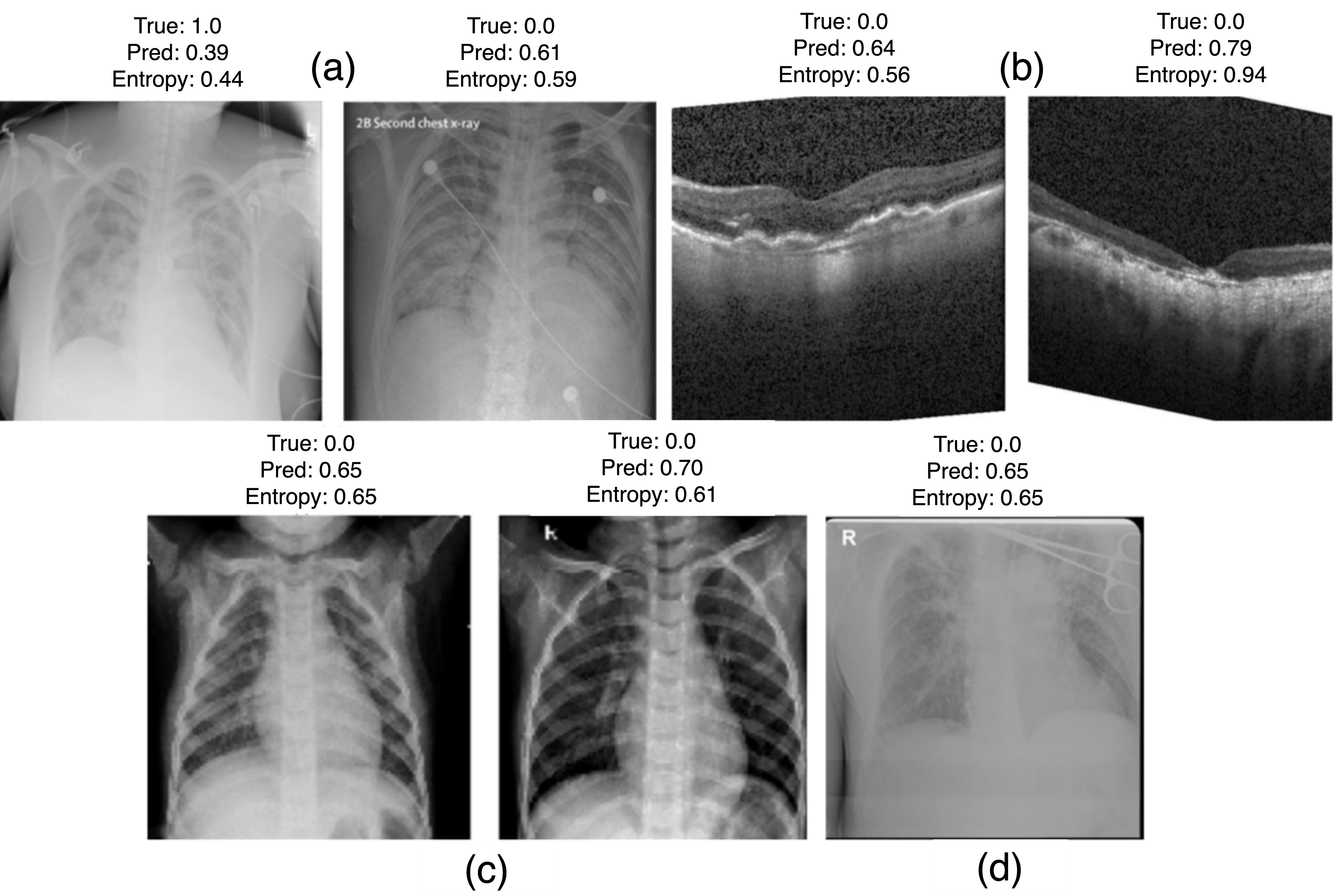}
    \caption{Relationship between entropy-based uncertainty and conformal prediction set size. Samples with higher entropy values produce larger prediction sets, indicating consistent reliability estimation between probabilistic and conformal approaches. The samples correspond to the following from the left : (a) Covid-19 Chest X-ray, (b) retinal OCT, (c) Pneumonia Chest X-ray, and (d) Tuberculosis Chest X-ray dataset.
    Images show high entropy-based uncertain samples that can be flagged for further investigation by the clinician.}\label{fig7}
\end{figure}

\begin{table}[!h]
\caption{Performance comparison of reported results in existing literature and proposed DCAT fusion model}\label{tab5}
\begin{tabular*}{\textwidth}{@{\extracolsep\fill}lcccccccc}
\toprule%
\textbf{Methods} & \textbf{Datasets} & \textbf{Acc} & \textbf{AUC} & \textbf{UQ} & \textbf{Mean Entropy} & \shortstack{\textbf{Standard}\\\textbf{deviation}}  & \textbf{HUS} & \textbf{CP}\\
\midrule

\cite{abdar2021barf}&Pneumonia  & 81.57 & 92.60 & Yes &  \quad-- &  \quad-- &  \quad-- & \ding{55}\\

\cite{yang2023medmnist} &Pneumonia  & 94.60 & 99.10 & No & \quad--   &  \quad-- &  \quad-- & \ding{55}\\

\cite{chhikara2020deep} &Pneumonia  & 90.70 & \quad-- & No & \quad--   &  \quad-- &  \quad-- & \ding{55}\\

\textbf{Ours} &Pneumonia & \textbf{98.95} & \textbf{99.93} & Yes & 0.0150 & 0.0817 & 13 & \ding{51}\\
\hline
\cite{abdar2021barf} &OCT & \textbf{91.40} & \textbf{98.21} & Yes & \quad-- & \quad-- &  \quad-- & \ding{55}\\

\cite{yang2023medmnist} &OCT  & 77.60 & 96.20 & No. & \quad--  & \quad-- &  \quad-- & \ding{55}\\

\textbf{Ours} &OCT & \textbf{91.17} & \textbf{98.60} & Yes & 0.1603  & 0.2457 & 45 & \ding{51}\\
\bottomrule
\end{tabular*}
\footnotetext{HUS: High Uncertainty Sample}
\end{table}

\begin{figure}[h!]
\centerline{\includegraphics[width=.98\columnwidth]{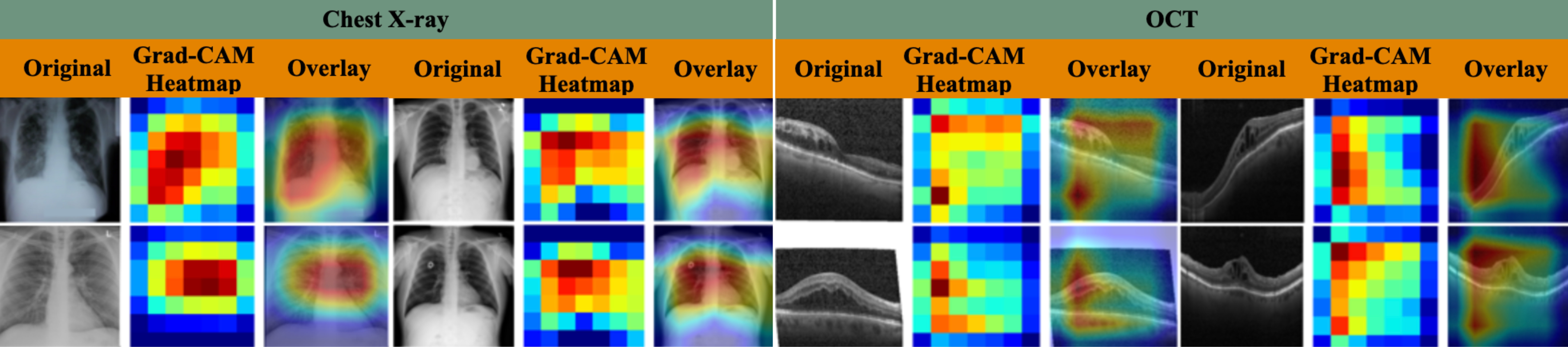}}
\caption{Visualization of Grad-CAM analysis on chest X-ray and OCT datasets using the proposed method from the left: Original chest X-ray and OCT images, their corresponding heat maps, and the Grad-CAM overlays. Heat maps and their Grad-CAM overlays with original images illustrate the regions that are of most importance for the network to lead to an interpretation of a clinical condition. Heat maps show relative activation intensity across regions of the image, where the warmer colors represent the higher activation values. Their Grad-CAM overlay images indicate the regions contributing strongly to the prediction of the network.}
\label{fig8}
\end{figure}

\begin{table}[h!]
\caption{Evaluation of CP across datasets. A 90\% confidence level is taken for empirical coverage, while a value close to 1 for average prediction set size remains close to 1, indicating the model is calibrated and confidently predicted on most samples}\label{tab6}
\centering
\begin{tabular}{lcccc}
\toprule
\textbf{Datasets} & \shortstack{\textbf{Classification}\\\textbf{type}} & \shortstack{\textbf{Confidence}\\\textbf{level $(1-\alpha)$}} & \shortstack{\textbf{Empirical}\\\textbf{Coverage}} & \shortstack{\textbf{Average Prediction}\\\textbf{Set Size}} \\
\hline
Covid-19 & Binary & 90\% & 89.27\%  &  1.12 \\

Tuberculosis & Binary & 90\%  & 96.88\% & 1  \\

Pneumonia & Binary & 90\% & 89.97\% & 1.07 \\

OCT & Multiclass & 90\% & 89.98\% & 1.06 \\
\bottomrule
\end{tabular}
\end{table}

\subsection{Discussion}\label{sec6}
In the last few decades, the impressive performance of DL methods has been broadly applied to various fields, expanding from engineering to healthcare in analyzing medical data. However, they are often criticized for their limitations in interpretability, resulting in challenges in their clinical adoption, which is essential for transparency and reliability. Inspired by the previous studies, in this work, interpretability and reliability are addressed as a combination of attention-based feature fusion representation with uncertainty-aware inference. This enables visual as well as quantitative assessment of model behavior. Uncertainty quantification with salient region visualization provides complementary interpretation beyond point predictions, supporting informed clinical decision-making. For qualitative assessment of interpretability, saliency maps are generated from heat maps and gradient-weighted class activation mapping (Grad-CAM) methods, which is illustrated in Figure~\ref{fig8}. This highlights the effectiveness of the proposed framework to produce more clinically meaningful and coherent feature localization while maintaining confidence in prediction.

To better reveal the importance and impact of our proposed framework, AUPR and MCC values are shown in Table~\ref{tab3}, and its performance is compared with previous methods in Table~\ref{tab5}. Finally, the probability distribution of the predictions for each class obtained with four datasets is presented in Table~\ref{tab2}. In Table~\ref{tab1}, we show some high-uncertainty samples (HUS) and misclassified samples. All uncertain samples above a threshold point are shown in Figure~\ref{fig6}, whereas \ref{fig7} demonstrates some of the highly uncertain samples with a large prediction set size from each dataset. These highly uncertain samples can be flagged for the medical experts for further investigation and clarification. Altogether, the key advantage of using the proposed framework is as follows: 1) hierarchical dual cross-attention, where each model contributes to the final prediction with an advantage of feature fusion closely attending and informing what to look at and where to look at, 2) an uncertainty quantification method to evaluate the prediction and uncertainties, giving the number of highly uncertain and misclassified samples, 3) because our datasets are class-imbalanced, the AUC, AUPR, and MCC values in Table~\ref{tab3} shows the confidence of our method in discriminating between positive and negative cases while making some errors. 

In medical imaging, translating a diagnosis into clinically actionable information, the evaluation of any health condition is guided by the physician's intuition and knowledge. In contrast, using AI-driven diagnostic tools, the reliability will be on the model's identification and understanding of features that align with medical practice-based evidence. Different variations in medical imaging can lead to variations in the DL model's performance, such as specific medical imaging modality, which has its unique challenges, data availability and tasks, resolution of images, and preprocessing requirements, information contents, architectural complexity within the model, and clinical variability. Considerations incorporating such a spectrum into the design of the model, training, and evaluation would be effective and transferable in clinical applications. Quantifying uncertainty within the model's predictions offers a more robust and scalable approach. This facilitates confidence in the interpretation of medical images for clinical decision-making, while contributing to the interpretability of the model. It allows clinicians to identify cases that are most uncertain for further investigation and verification. 

\section{Conclusion}\label{sec7}
This study proposes a simple yet effective feature fusion framework with uncertainty quantification, based on a hierarchical dual cross-attention fusion approach that combines two pretrained deep learning models. The proposed framework was evaluated on well-known medical chest X-ray and OCT datasets, demonstrating strong classification performance. The framework effectively captured spatial and contextual feature representations. The approach framed the attention as a selective, hierarchical feature fusion mechanism and learned multi-scale feature representations. We plan to extend our proposed framework to the medical image segmentation task and further enhance uncertainty modeling to achieve adaptive and robust performance.

\section*{Funding}
The authors declare that no funds, grants, or other support were received during the preparation of this manuscript.

\section*{Data Availability}
All the information about the data is included in the manuscript and is available publicly.

\section*{Competing Interests}
The authors declare no competing interests.

\section*{Author Contributions}
All authors contributed to the study conception and design. Methodology preparation, experimentation, and analysis were performed by Jutika Borah. The first draft of the manuscript was written by Jutika Borah. Revisions and comments on the manuscript were performed and provided by Hidam Kumarjit Singh. All authors read and approved the final manuscript.

\section*{Ethics approval}
This is an observational study. So, no ethical approval is required.

\section*{Consent to participate}
Not applicable.

\section*{Consent to publish}
This manuscript has not been published or presented elsewhere in part or in entirety and is not under consideration by another journal. We have read and understood your journal’s policies, and we believe that neither the manuscript nor the study violates any of these.

\bibliography{sn-bibliography}

\end{document}